\documentclass[12pt]{article}
\usepackage{amssymb}
\usepackage{graphicx}
\usepackage{amsmath}

\input{tcilatex}

\begin{document}

\begin{center}
{\LARGE Path-Free Decomposition for Direct, Indirect and Interaction Effects
in Mediation Analysis}

(October 2021)

\begin{tabular}{c}
Myoung-jae Lee \\ 
Department of Economics \\ 
Korea University \\ 
145 Anam-ro, Sungbuk-gu \\ 
Seoul 02841, South Korea \\ 
myoungjae@korea.ac.kr \\ 
phone/fax:\ 82-2-3290-2229%
\end{tabular}%
\bigskip \bigskip \bigskip \bigskip \bigskip \bigskip \bigskip
\end{center}

Given a binary treatment and a binary mediator, mediation analysis
decomposes the total effect of the treatment on an outcome variable into
direct and indirect effects. However, the existing decompositions are
\textquotedblleft path-dependent\textquotedblright , and consequently, there
appeared different versions of direct and indirect effects. Differently from
these, this paper proposes a \textquotedblleft path-free\textquotedblright\
decomposition of the total effect into three sub-effects: direct, indirect,
and treatment-mediator interaction effects. Whereas the interaction effect
has been part of the indirect effect in the existing two-effect
decompositions, it is separately identified in our three-effect
decomposition. All effects are found using conditional means, but not
conditional densities, and are estimated with ordinary least squares
estimators. Simulation and empirical analyses are provided as well.\bigskip
\bigskip \bigskip \bigskip 

\medskip

\textbf{Compliance with ethical standard \& no conflict of interest:}\ no
human or animal subject is involved in this research, and there is no
conflict of interest to disclose.\pagebreak

\section{Introduction}

\qquad Given a binary treatment $D$, a binary mediator $M$ and an
outcome/response variable $Y$, the causal chain of interest in mediation
analysis is%
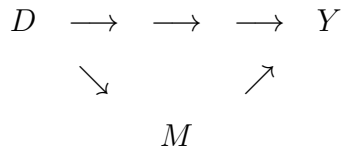
\begin{equation*}
\begin{tabular}{ccccc}
$D$ & $\longrightarrow $ & $\longrightarrow $ & $\longrightarrow $ & $Y$ \\ 
& $\searrow $ &  & $\nearrow $ &  \\ 
&  & $M$ &  & 
\end{tabular}%
\end{equation*}%
where the total effect of $D$ on $Y$ consists of the direct effect of $D$ on 
$Y$ and the indirect effect of $D$ on $Y$ through $M$. This is an important
issue in various disciplines of science, as reviewed in MacKinnon et al.
(2007), Pearl (2009), Imai et al. (2010), TenHave and Joffe (2012), Preacher
(2015), VanderWeele (2015, 2016) and Nguyen et al. (2021).

\qquad Finding the total effect of $D$ on $Y$ can be done in various ways
such as matching, regression adjustment, weighting, etc. Traditionally,
decomposing the total effect has been done straightforwardly, using linear
structural-form (SF) models for $Y$ as a function of $(D,M)$ and $M$ as a
function of $D$ (unless the interaction term $DM$ appears in the $Y$ SF).
However, this type of traditional approaches are model-dependent, and not
exactly causal from the viewpoint of modern causal analysis. Once we leave
linear SF's for nonparametric approaches to avoid misspecifications while
introducing potential variables for $(M,Y)$, decomposing the total effect is
no more straightforward.

\qquad Consider two potential versions $M^{d}$, $d=0,1$, of $M$
corresponding to $D=0,1$, and the four potential responses $Y^{dm}$ for $D=d$
and $M=m$ with $d,m=0,1$. Also define the potential responses $Y_{d}$, $%
d=0,1 $, corresponding to $D=0,1$ \textquotedblleft when $M$ is allowed to
take its natural course given $D=d$\textquotedblright :%
\begin{equation*}
Y_{d}\equiv Y^{d,M^{d}}.
\end{equation*}%
Then we have the `total effect' $\tau $:%
\begin{equation*}
\text{\textit{total effect}}:\tau \equiv Y_{1}-Y_{0}=Y^{1,M^{1}}-Y^{0,M^{0}}.
\end{equation*}

\qquad The `natural direct effect' of Pearl (2001) is ($\delta $ in $\delta
(d)$ is from d for `direct'):%
\begin{eqnarray}
&&\ \ \text{\textit{natural direct effect with }}M^{d}\ :\ \delta (d)\equiv
Y^{1,M^{d}}-Y^{0,M^{d}}  \TCItag{1.1} \\
&&\ \ \ \Longrightarrow \ \delta (0)\equiv Y^{1,M^{0}}-Y^{0,M^{0}},\ \ \
\delta (1)\equiv Y^{1,M^{1}}-Y^{0,M^{1}};  \notag
\end{eqnarray}%
Robins (2003) called $\delta (d)$ the `pure or total direct effect'. The
`natural indirect effect' of Pearl (2001) is ($\mu $ in $\mu (d)$ is from m
in `mediator'):%
\begin{eqnarray}
&&\ \ \text{\textit{natural indirect effect with }}d\ :\ \mu (d)\equiv
Y^{d,M^{1}}-Y^{d,M^{0}}  \TCItag{1.2} \\
&&\ \ \ \Longrightarrow \ \mu (0)\equiv Y^{0,M^{1}}-Y^{0,M^{0}},\ \ \ \mu
(1)\equiv Y^{1,M^{1}}-Y^{1,M^{0}};  \notag
\end{eqnarray}%
Robins (2003) called $\mu (d)$ the `pure or total indirect' effect.

\qquad These effect identification and estimation have been addressed by
Pearl (2001), Robins (2003), Peterson (2006) and Tchetgen Tchetgen and
Shpitser (2012, 2014), among others. Differently from the natural effects,
however, central to our paper are%
\begin{eqnarray}
\text{\textit{controlled direct effect with }}m &:&Y^{1,m}-Y^{0,m},  \notag
\\
\text{\textit{controlled mediator effect with} }d &:&Y^{d,1}-Y^{d,0}; 
\TCItag{1.3}
\end{eqnarray}%
the name `controlled mediator effect' is adopted from TenHave and Joffe
(2012).

\qquad The two well-known ways to decompose the total effect $\tau $ is%
\begin{eqnarray}
\tau &=&Y^{1,M^{1}}-Y^{1,M^{0}}+Y^{1,M^{0}}-Y^{0,M^{0}}=\mu (1)+\delta (0), 
\TCItag{1.4} \\
\tau &=&Y^{1,M^{1}}-Y^{0,M^{1}}+Y^{0,M^{1}}-Y^{0,M^{0}}=\delta (1)+\mu (0). 
\TCItag{1.5}
\end{eqnarray}%
These two decompositions can be written succinctly as%
\begin{equation}
\tau =\mu (d)+\delta (1-d)\text{ \ \ \ \ for \ \ }d=0,1,  \tag{1.6}
\end{equation}%
which is, however, `$d$- or path-dependent'. This is one problem, and
another problem is that some effects are relative to $d=0$ while some others
are relative to $d=1$; e.g., $\delta (0)$ in (1.4) is for the change in $D$
relative to the untreated mediator $M^{0}$, but $\mu (1)$ in (1.4) is for
the change in $M$ relative to $d=1$, not to $d=0$. Hence, Pearl (2009, p.
132) even stated \textquotedblleft the total effect \textit{TE} of a
transition is equal to the \textit{difference} between the direct effect of
that transition and the indirect effect of the reverse
transition\textquotedblright .

\qquad There are many generalizations of (1.6), and we could not cover
possibly all of them here. Just to mention a few, effects of `stochastic
interventions' on $D$ or $M$ (i.e., `interventional effects') appeared in
VanderWeele et al. (2014), Lok (2016), Vansteelandt and Daniel (2017),
VanderWeele and Tchetgen Tchetgen (2017), Diaz and Hejazi (2020), Diaz et
al. (2021) and Nguyen et al. (2021), among others. Instruments for $D$ were
considered in mediation analysis, e.g., by Joffe et al. (2008), Fr\"{o}lich
and Huber (2017) and Rudolph et al. (2021). Forastiere et al. (2018) adopted
principal stratifications for mediation analysis, building on Rubin (2004),
Jo and Stuart (2009) and Ding and Lu (2017), which are related to our
approach because $M^{0}\leq M^{1}$ will be invoked sometimes.

\qquad In (1.4) and (1.5), decomposing $\tau $ (or $E(\tau )$) into direct
and indirect effects was done by subtracting and adding a \textquotedblleft
cross-world\textquotedblright\ potential outcome such as $Y^{1,M^{0}}$ or $%
Y^{0,M^{1}}$, and depending on which was used, different decompositions were
obtained, which is a path-dependence. This is an important issue, and in a
nut shell, the goal of this paper is to propose a path-independent or
`path-free' decomposition of $\tau $ or $E(\tau )$.

\qquad Our approach has two advantages compared with the existing
approaches: the first is the aforementioned path independence, and the
second is that we obtain a more informative three-effect decomposition, not
two as in the existing approaches. The extra effect is the effect of the
interaction term $DM$ on $Y$, which has been buried in the indirect effect
in the existing decompositions. Our approach also has a limitation: only a
single binary $D$ and a single binary $M$ are allowed, neither multi-valued
nor multiple treatments or mediators. Nevertheless, binary $D$ and $M$ are
building blocks for more general $D$ and $M$, to which our approach may get
extended in the future.

\qquad In the remainder of this paper, Section 2 introduces our path-free
three-effect decompositions. Section 3 explains how the effects in our
decomposition can be identified. Section 4 examines simple linear SF's for $%
M $ and $Y$ to exemplify what the effects actually look like. Section 5
introduces two estimators for our three-effect decompositions. Section 6
conducts a simulation study to show that the estimators work well, and then
provides an empirical analysis. Finally, Section 7 concludes this paper.

\section{Path-Free Decomposition of Total Effect}

\qquad For our path-free decomposition, the first step is rewriting $Y_{0}$
and $Y_{1}$:%
\begin{equation}
Y_{0}\equiv Y^{0,M^{0}}=Y^{00}+(Y^{01}-Y^{00})M^{0},\text{ \ \ }Y_{1}\equiv
Y^{1,M^{1}}=Y^{10}+(Y^{11}-Y^{10})M^{1};  \tag{2.1}
\end{equation}%
the equalities can be seen by substituting $M^{0}=0,1$ and $M^{1}=0,1$.

\subsection{Basic Three-Effect Decomposition}

\qquad With (2.1), the total effect $E(\tau )\equiv E(Y_{1}-Y_{0})$ can be
decomposed \textquotedblleft path-freely\textquotedblright :%
\begin{eqnarray}
&&E(Y_{1}-Y_{0})=E[\
Y^{10}+(Y^{11}-Y^{10})M^{1}-\{Y^{00}+(Y^{01}-Y^{00})M^{0}\}\ ]  \notag \\
&=&E(Y^{10}-Y^{00})\ +\ E\{(Y^{11}-Y^{10})M^{1}\}-E\{(Y^{01}-Y^{00})M^{0}\}.
\TCItag{2.2}
\end{eqnarray}%
We call $(M^{0}=0,M^{1}=0)$ `never taker', $(M^{0}=0,M^{1}=1)$ `complier', $%
(M^{0}=1,M^{1}=0)$ `defier', and $(M^{0}=1,M^{1}=1)$ `always taker'. These
terms were used in Imbens and Angrist (1994) when $M$ is an endogenous
treatment and $D$ is an instrument, but we use those terms for mediator $M$
and treatment $D$ as in Lee (2012, 2017) where $M$ is participation in an
activity and $Y$ is a performance in the activity.

\qquad Now, subtract and add $E\{(Y^{11}-Y^{10})M^{0}\}$ to (2.2) to rewrite
(2.2) as%
\begin{equation}
E(Y^{10}-Y^{00})+E\{(Y^{11}-Y^{10})(M^{1}-M^{0})\}+E%
\{(Y^{11}-Y^{10}-Y^{01}+Y^{00})M^{0}\}.  \tag{2.3}
\end{equation}%
\textit{The first term is the controlled direct effect with }$m=0$\textit{\
in (1.3). The second term can be called the `controlled indirect effect',
because }$M^{1}-M^{0}$\textit{\ is the effect of }$D$\textit{\ on }$M$%
\textit{\ and }$Y^{11}-Y^{10}$\textit{\ is the controlled mediator effect
with }$d=1$\textit{\ in (1.3). The third term is the `controlled interaction
effect' of }$D$\textit{\ and }$M$, as is explained below.

\qquad To relate the indirect effect in (2.3) to the traditional `product
approach', consider%
\begin{equation*}
M=\alpha _{1}+\alpha _{d}D+\varepsilon \text{ \ \ \ \ and \ \ \ \ }Y=\beta
_{1}+\beta _{d}D+\beta _{m}M+U
\end{equation*}%
where the $\alpha $'s and $\beta $'s are parameters, $\varepsilon $ and $U$
are error terms. Here, the effect of $D$ on $M$ is $\alpha _{d}$, the effect
of $M$ on $Y$ is $\beta _{m}$, and consequently, the indirect effect of $D$
on $Y$ through $M$ is $\alpha _{d}\beta _{m}$. The second term of (2.3) is a
nonparametric version of $\beta _{m}\alpha _{d}$.

\qquad To understand the third term of (2.3) intuitively, consider $Y=\beta
_{1}+\beta _{d}D+\beta _{m}M+\beta _{dm}DM+U$ with $\beta _{dm}DM$ extra,
compared with the preceding SF for $Y$. Observe%
\begin{equation*}
Y^{11}-Y^{10}-Y^{01}+Y^{00}=Y^{11}-Y^{00}-(Y^{10}-Y^{00})-(Y^{01}-Y^{00})
\end{equation*}%
which is the `gross effect ($\beta _{d}+\beta _{m}+\beta _{dm}$) $%
Y^{11}-Y^{00}$ of $(D,M)$' minus the `separate effect ($\beta _{d}$) $%
Y^{10}-Y^{00}$ of $D$' minus the `separate effect ($\beta _{m}$) $%
Y^{01}-Y^{00}$ of $M$'. That is, the double difference in the third term of
(2.3) removes the separate effects ($\beta _{d}$ and $\beta _{m}$) of $D$
and $M$ from the gross effect to isolate only their interaction effect $%
\beta _{dm}$.

\subsection{Three-Effect Decomposition under Monotonicity}

\qquad To better interpret (2.3), we can rule out defier under the assumption%
\begin{equation*}
\text{Monotonicity}:M^{0}\leq M^{1}\text{\ }\Longrightarrow \
M^{1}-M^{0}=0,1.
\end{equation*}%
This makes the indirect and interaction effects of (2.3) equal to,
respectively,%
\begin{eqnarray*}
&&E(Y^{11}-Y^{10}|M^{1}-M^{0}=1)P(M^{1}-M^{0}=1)=E(Y^{11}-Y^{10}|\text{CP})P(%
\text{CP}), \\
&&E(Y^{11}-Y^{10}-Y^{01}+Y^{00}|\text{AT})P(\text{AT}),
\end{eqnarray*}%
where `CP' and `AT' are shorthands for complier and always taker. Then (2.3)
becomes%
\begin{eqnarray}
&&E(Y^{10}-Y^{00})+E(Y^{11}-Y^{10}|\text{CP})P(\text{CP}%
)+E(Y^{11}-Y^{10}-Y^{01}+Y^{00}|\text{AT})P(\text{AT})  \notag \\
&&\ =\text{direct effect}+\text{mediator effect for CP}+\text{interaction
effect for AT;}  \TCItag{2.4}
\end{eqnarray}%
being CP represents the effect of $D$ on $M$, as will become clear later.

\qquad We summarize our findings for the three-effect decomposition:\bigskip

\textbf{THEOREM 1. }\textit{A path-free three-effect decomposition of the
total effect }$E(Y_{1}-Y_{0})$\textit{\ is (2.3), consisting of (i) the
controlled direct effect }$E(Y^{10}-Y^{00})$\textit{, (ii) the controlled
indirect effect }$E\{(Y^{11}-Y^{10})(M^{1}-M^{0})\}$\textit{, and (iii) the
controlled interaction effect }$E\{(Y^{11}-Y^{10}-Y^{01}+Y^{00})M^{0}\}$%
\textit{. If }$M^{0}\leq M^{1}$\textit{\ holds extra, then (2.3) becomes
(2.4).\bigskip }

\qquad In (2.4), if $E(Y^{11}-Y^{10}|$CP$)=0$ or $P($CP$)=0$, then the
second term is zero. If $E(Y^{11}-Y^{10}-Y^{01}+Y^{00}|$AT$)=0$ or $P($AT$%
)=0 $, then the third term is zero. The fact that only the compliers appear
in the controlled mediator effect under the monotonicity is natural, because
the defiers ($M^{1}=0$ and $M^{0}=1$) are ruled out and $M$ changes neither
(i.e., $M^{0}=M^{1}$) for always takers nor for never takers.

\qquad One may define the `controlled total effect' $E(Y^{11}-Y^{00})$, and
decompose it as in%
\begin{eqnarray}
E(Y^{11}-Y^{00}) &=&E(Y^{11}-Y^{10})+E(Y^{10}-Y^{00}),  \TCItag{2.5} \\
E(Y^{11}-Y^{00}) &=&E(Y^{11}-Y^{01})+E(Y^{01}-Y^{00}).  \TCItag{2.6}
\end{eqnarray}%
Although these are similar to (2.3) in that they are based on controlled
effects, there are two critical differences. The first is that $%
E(Y^{11}-Y^{00})$ differs from the total effect $E(\tau )\equiv
E(Y_{1}-Y_{0})$ for (2.3), and $E(Y^{11}-Y^{00})$ is not germane to
mediation analysis, because both $D$ and $M$ are controlled as two causal
factors of equal standing with none preceding the other. The second is that
(2.3) is path-free,\ while (2.5) and (2.6) are not.

\section{Identification of All Effects}

\subsection{Identification Conditions}

\qquad Let $X$ be covariates not affected by $D$. Our conditions are (`$%
\amalg $' for independence):\bigskip

\textbf{C(a) :} $D\amalg (M^{0},M^{1},Y^{00},Y^{01},Y^{10},Y^{11})|X;$

\textbf{C(b) :\ }$(M^{0},M^{1})\amalg (Y^{00},Y^{01},Y^{10},Y^{11})|(D,X);$

\textbf{C(c) : }$0<P(D=d,M=m|X)$ for all $d,m=0,1$ and $X;$

\textbf{C(d)\ : }$M^{0}\leq M^{1}|X.\bigskip $

C(a) to C(c) are essential for (2.3). C(d) is for (2.4), which is not
essential.

\qquad C(a) and C(b) are the ignorability of confounders in the
treatment-mediator, treatment-outcome, and mediator-outcome relationships.
C(a) and C(b) operate in two stages: as $D$ precedes $M$ which in turn
precedes $Y$, the first stage is $D$ being independent of all potential
future\ variables given $X$, and the second stage is $(M^{0},M^{1})$ being
independent of all potential future\ versions of $Y$ given $(D,X)$. C(c) is
a support-overlap condition for $D|X$, $D|(M,X)$ and $M|(D,X)$; e.g., C(c)
implies%
\begin{eqnarray*}
P(D &=&d|X)=P(D=d,M=0|X)+P(D=d,M=1|X)>0\text{ \ \ for \ \ }d=0,1; \\
P(M &=&m|D=d,X)=P(D=d,M=m|X)/P(D=d|X)>0\text{ \ \ for }\ \ d,m=0,1.
\end{eqnarray*}

\qquad Slightly weaker conditions than C(a) and C(b) appeared in Imai et al.
(2010):%
\begin{equation*}
D\amalg (M^{d},Y^{d^{\prime }m})|X\text{ \ \ and \ \ }M^{d}\amalg
Y^{d^{\prime }m}|(D,X)\text{ \ \ for all }d,d^{\prime },m=0,1
\end{equation*}%
where the joint distributions of $(M^{0},M^{1})$ and $%
(Y^{00},Y^{01},Y^{10},Y^{11})$ do not appear, differently from C(a) and
C(b). Analogously, Petersen et al. (2006) assumed%
\begin{equation*}
D\amalg M^{d}|X,\ \ \ D\amalg Y^{dm}|X,\ \ \ M\amalg Y^{dm}|(D,X)\ \text{\ \
for all }d,m=0,1.
\end{equation*}%
Using marginal independence instead of joint independence, we can relax C(a)
and C(b), but we continue to assume C(a) and C(b) for simplicity.

\qquad In C(b), $(M^{0},M^{1})$ is allowed to be related to $%
(Y^{00},Y^{01},Y^{10},Y^{11})$ through $(D,X)$, but $D\amalg
(M^{0},M^{1},Y^{00},Y^{01},Y^{10},Y^{11})|X$ in C(a). Hence, \textit{C(a)
and C(b) imply C(e) next.}\bigskip

\textbf{C(e)}\ : $(D,M^{0},M^{1})\amalg (Y^{00},Y^{01},Y^{10},Y^{11})|X\
\{\Longrightarrow \ (D,M)\amalg (Y^{00},Y^{01},Y^{10},Y^{11})|X\};$\bigskip

the implication arrow holds as $M=M^{0}+(M^{1}-M^{0})D$ is determined by $%
(M^{0},M^{1},D)$.

\subsection{Effect Identification}

\qquad First, the total effect $E(\tau )\equiv E(Y_{1}-Y_{0}|X)$ is easily
identified:%
\begin{eqnarray}
&&E(Y|D=1,X)-E(Y|D=0,X)=E(Y_{1}|D=1,X)-E(Y_{0}|D=0,X)  \TCItag{3.1} \\
&=&E\{Y^{10}+(Y^{11}-Y^{10})M^{1}|D=1,X\}-E%
\{Y^{00}+(Y^{01}-Y^{00})M^{0}|D=0,X\}  \notag \\
&=&E\{Y^{10}+(Y^{11}-Y^{10})M^{1}|X\}-E\{Y^{00}+(Y^{01}-Y^{00})M^{0}|X%
\}=E(Y_{1}-Y_{0}|X);  \notag
\end{eqnarray}%
the third equality is due to C(a). This is sensible, as only $D$ changes
from $0$ to $1$ in (3.1).

\qquad Second, the direct effect $E(Y^{10}-Y^{00}|X)$ of $D$ with $m=0$ is
also easily identified:%
\begin{eqnarray}
&&E(Y|D=1,M=0,X)-E(Y|D=0,M=0,X)  \TCItag{3.2} \\
&=&E(Y^{10}|D=1,M^{1}=0,X)-E(Y^{00}|D=0,M^{0}=0,X)=E(Y^{10}-Y^{00}|X)  \notag
\end{eqnarray}%
due to C(e). This is sensible, as only $D$ changes from $0$ to $1$ with $M=0$%
.

\qquad Third, due to C(e), the indirect effect given $X$ can be written as%
\begin{eqnarray}
&&E\{(Y^{11}-Y^{10})(M^{1}-M^{0})|X\}=E(Y^{11}-Y^{10}|X)\cdot
E(M^{1}-M^{0}|X)  \notag \\
&=&\{E(Y|D=1,M=1,X)-E(Y|D=1,M=0,X)\}  \notag \\
&&\ \cdot \{E(M|D=1,X)-E(M|D=0,X)\}  \TCItag{3.3}
\end{eqnarray}%
because, respectively due to C(e) and C(a),%
\begin{eqnarray}
&&E(Y|D=1,M=1,X)-E(Y|D=1,M=0,X)=E(Y^{11}-Y^{10}|X);  \TCItag{3.4} \\
&&E(M|D=1,X)-E(M|D=0,X)=E(M^{1}|X)-E(M^{0}|X)=E(M^{1}-M^{0}|X).  \notag
\end{eqnarray}

\qquad Fourth, the interaction effect is identified as the remainder $%
(3.1)-(3.2)-(3.3)$.

\qquad Now, under C(d), apply C(e) to (3.3) to get the effect of $M$ with $%
d=1$ on compliers:%
\begin{eqnarray}
&&E(Y|D=1,M=1,X)-E(Y|D=1,M=0,X)=E(Y^{11}-Y^{10}|X)  \notag \\
&=&E(Y^{11}-Y^{10}|M^{1}-M^{0}=1,X)=E(Y^{11}-Y^{10}|\text{CP},X). 
\TCItag{3.5}
\end{eqnarray}%
Also in (3.3), due to C(d) and C(a),%
\begin{eqnarray}
&&\text{`effect of }D\text{ on }M\text{'}=E(M|D=1,X)-E(M|D=0,X)  \TCItag{3.6}
\\
&&\ =E(M^{1}|X)-E(M^{0}|X)=P(M^{1}=1,M^{0}=1|X)+P(M^{1}=1,M^{0}=0|,X)  \notag
\\
&&\ -P(M^{1}=1,M^{0}=1|X)=P(M^{1}=1,M^{0}=0|X)=P(\text{CP}|X).  \notag
\end{eqnarray}

\qquad The next theorem summarizes the main findings for effect
identification:\bigskip

\textbf{THEOREM 2.} \textit{Under C(a) and C(c), the identification findings
are:}%
\begin{eqnarray*}
&&\text{\textit{total effect}}:E(Y|D=1,X)-E(Y|D=0,X); \\
&&\text{\textit{controlled direct}}:E(Y|D=1,M=0,X)-E(Y|D=0,M=0,X); \\
&&\text{\textit{controlled indirect}}:\{E(Y|D=1,M=1,X)-E(Y|D=1,M=0,X)\} \\
&&\ \ \ \ \ \ \ \ \ \ \ \ \ \ \ \ \ \ \ \ \ \ \ \ \ \ \cdot
\{E(M|D=1,X)-E(M|D=0,X)\}\text{ \ \ \ \ \textit{(for CP under C(d))}}; \\
&&\text{\textit{controlled interaction}}:\text{\textit{total}}-\text{\textit{%
controlled direct }}-\text{ \textit{controlled indirect.}}
\end{eqnarray*}

\section{Effect Comparison in Linear Model}

\subsection{Structural Form and Reduced Form}

\qquad To understand various effect decompositions and identifications
better, here we illustrate the effects using a randomized $D$ with $%
P(D=1)=0.5$ and linear SF's:%
\begin{eqnarray}
M^{d} &=&1[1<\alpha _{1}+\alpha _{d}d+X^{\prime }\alpha _{x}+e]\text{,} 
\TCItag{4.1} \\
Y^{dm} &=&\beta _{1}+\beta _{d}d+\beta _{m}m+\beta _{dm}dm+X^{\prime }\beta
_{x}+U,  \notag
\end{eqnarray}%
where $1[A]\equiv 1$ if $A$ holds and $0$ otherwise, the error terms $(e,U)$
are independent of each other and $X$, and $e\sim Uni[0,1]$ with $Uni[0,1]$
standing for the uniform distribution on $[0,1]$. The reason for $Uni[0,1]$
and the threshold $1$ for $M^{d}$ is to have a linear model for $E(M^{d}|X)$%
, as is shown next.

\qquad Assuming $0<\alpha _{1}+\alpha _{d}d+X^{\prime }\alpha _{x}<1$ for
all $X$, it holds that%
\begin{eqnarray}
&&E(M^{d}|X)=P(e>1-\alpha _{1}-\alpha _{d}d-X^{\prime }\alpha
_{x}|X)=P(e<\alpha _{1}+\alpha _{d}d+X^{\prime }\alpha _{x}|X)  \notag \\
&&\ =\alpha _{1}+\alpha _{d}d+X^{\prime }\alpha _{x}\ \Longrightarrow \
M^{d}=\alpha _{1}+\alpha _{d}d+X^{\prime }\alpha _{x}+\varepsilon ^{d} 
\TCItag{4.2} \\
&&\ \ \ \text{where }\ \ \ \ \varepsilon ^{d}\equiv M^{d}-\alpha _{1}-\alpha
_{d}d-X^{\prime }\alpha _{x}\text{ }\{\Longrightarrow \text{\ }E(\varepsilon
^{d}|X)=0\}.  \notag
\end{eqnarray}

$M^{d}=\alpha _{1}+\alpha _{d}d+X^{\prime }\alpha _{x}+\varepsilon ^{d}$ is
a reduced form (RF) in contrast to the SF for $M^{d}$ in (4.1), and $%
E(M^{d}|X)=\alpha _{1}+\alpha _{d}d+X^{\prime }\alpha _{x}$ will be used
often. `$e\sim Uni[0,1]$' is restrictive, but not much more restrictive than
the probit assumption as the following shows.

\qquad Consider a $M^{d}$ SF with an error $\nu $ having an invertible
distribution function $F$:%
\begin{equation}
M^{d}=1[\nu >F^{-1}(1-\alpha _{1}-\alpha _{d}d-X^{\prime }\alpha
_{x})]=1[F(\nu )>1-\alpha _{1}-\alpha _{d}d-X^{\prime }\alpha _{x}]. 
\tag{4.3}
\end{equation}%
Since $F(\nu )\sim Uni[0,1]$ as $e$ is, (4.3) is restrictive only in that
the regression function part is assumed to be $F^{-1}(1-\alpha _{1}-\alpha
_{d}d-X^{\prime }\alpha _{x})$ when $\nu $ is the error term. Compare this
to the probit assumption:\ with the $N(0,1)$ distribution function $\Phi
(\cdot )$,%
\begin{equation}
M^{d}=1[N(0,1)>1-\alpha _{1}-\alpha _{d}d-X^{\prime }\alpha
_{x}]=1[Uni[0,1]>\Phi (1-\alpha _{1}-\alpha _{d}d-X^{\prime }\alpha _{x})], 
\tag{4.4}
\end{equation}%
which is almost as strong an assumption as (4.3) is.

\subsection{Various Effects for Linear Model}

\qquad The linear SF for $Y^{dm}$ in (4.1) and the linear RF for $M^{d}$ in
(4.2) render%
\begin{eqnarray}
&&\delta (d)=\beta _{d}+\beta _{dm}M^{d},\ \ \ \text{\ \ }\mu (d)=\beta
_{m}(M^{1}-M^{0})+\beta _{dm}d(M^{1}-M^{0})  \TCItag{4.5} \\
&&\ \Longrightarrow \ E(\tau )=\beta _{d}+\beta _{m}\alpha _{d}+\beta
_{dm}\{\alpha _{1}+\alpha _{d}+E(X^{\prime })\alpha _{x}\}\text{;} 
\TCItag{4.6}
\end{eqnarray}%
the proof is in the appendix. In contrast, the three-effect decomposition
(2.3) is%
\begin{eqnarray}
&&E\{Y^{10}-Y^{00}\ +(Y^{11}-Y^{10})(M^{1}-M^{0})\
+(Y^{11}-Y^{10}-Y^{01}+Y^{00})M^{0}\}  \notag \\
&&\ =E\{\beta _{d}\ +(\beta _{m}+\beta _{dm})(\alpha _{d}+\varepsilon
^{1}-\varepsilon ^{0})\ +\beta _{dm}(\alpha _{1}+X^{\prime }\alpha
_{x}+\varepsilon ^{0})\}  \notag \\
&&\ =\beta _{d}\ +\ (\beta _{m}+\beta _{dm})\alpha _{d}\ +\ \beta
_{dm}\{\alpha _{1}+E(X^{\prime })\alpha _{x}\}.  \TCItag{4.7}
\end{eqnarray}

\qquad The two-effect decomposition in (4.6) takes $\beta _{m}\alpha
_{d}+\beta _{dm}\{\alpha _{1}+\alpha _{d}+E(X^{\prime })\alpha _{x}\}$ as
the indirect effect. In contrast, our three-effect decomposition (4.7) takes
only $\beta _{m}\alpha _{d}+\beta _{dm}\alpha _{d}$ as the indirect effect
while classifying $\beta _{dm}\{\alpha _{1}+E(X^{\prime })\alpha _{x}\}$ as
the interaction effect, where $\beta _{dm}=E(Y^{11}-Y^{10}-Y^{01}+Y^{00})$
and $\alpha _{1}+E(X^{\prime })\alpha _{x}=P(M^{0}=1)$ are for%
\begin{equation*}
E\{(Y^{11}-Y^{10}-Y^{01}+Y^{00})M^{0}\}=E(Y^{11}-Y^{10}-Y^{01}+Y^{00})\cdot
P(M^{0}=1).
\end{equation*}%
Under $\beta _{dm}=0$ (no interaction effect), both (4.6) and (4.7) become $%
\beta _{d}+\beta _{m}\alpha _{d}$ which is the \textquotedblleft traditional
decomposition\textquotedblright\ of the total effect into the direct effect $%
\beta _{d}$ and the indirect effect $\beta _{m}\alpha _{d}$. Although $%
\alpha _{1}+E(X^{\prime })\alpha _{x}$ is irrelevant in the traditional
decomposition, they do matter in (4.6) and (4.7). Intuitively explaining
interaction effect only with $\beta _{dm}$ as was done earlier is not
exactly correct, as the interaction effect in (4.7) reveals.

\qquad We also mentioned other decompositions in (2.5) and (2.6), which are,
respectively,%
\begin{eqnarray}
\{\beta _{1}+\beta _{d}+\beta _{m}+\beta _{dm}-(\beta _{1}+\beta
_{d})\}+(\beta _{1}+\beta _{d}-\beta _{1}) &=&\beta _{m}+\beta _{dm}+\beta
_{d},  \notag \\
\{\beta _{1}+\beta _{d}+\beta _{m}+\beta _{dm}-(\beta _{1}+\beta
_{m})\}+(\beta _{1}+\beta _{m}-\beta _{1}) &=&\beta _{d}+\beta _{dm}+\beta
_{m}.  \TCItag{4.8}
\end{eqnarray}%
Whereas all the preceding decompositions involve the $\alpha $ parameters to
reveal how $M$ is affected by $D$, no $\alpha $ parameter appears in (4.8).
Since $M$ is affected by $D$ in reality, the decompositions in (4.8) without
any $\alpha $ parameter would make sense only when we control $M$ as well as 
$D$, which is, however, not a mediation analysis.

\section{Two Estimators}

\qquad There are many ways to estimate various effects identified with
conditional mean differences. The arguably best-known approaches are
matching, inverse probability weighting (IPW), and regression adjustment.
Among these, matching is most intuitive, but finding its standard error is
hard despite advances in Abadie and Imbens (2016). IPW specifies only $%
E(D|X) $, but it has the \textquotedblleft too small
denominator\textquotedblright\ problem, which remains even when IPW is
generalized for `doubly robustness'. For our goal, regression adjustment
specifying outcome models as in Vansteelandt and Daniel (2014) is well
suited.

\qquad In regression adjustment for the effect of $D$ on $Y$, $E(Y|D,X)$ is
specified to render the mean difference $E(Y|D=1,X)-E(Y|D=0,X)$, from which $%
X$ is averaged out. If the effect is constant, then the averaging step is
unnecessary. Hence we explore two estimators. The first is based on the
constant-effect linear models (4.1) and (4.2), the main attraction of which
is its simplicity. Even when the effect is heterogeneous, the misspecified
constant-effect models tend to give weighted versions of the heterogeneous
effect. The second estimator essentially replaces the constant effect
specifications with functions of $X$, and it relaxes the uniform error
assumption for $M^{d}$ in (4.1) and allows almost any form of $Y$ as is
explained next.

\qquad Generalizing the approaches in Lee (2018,\ 2021) without $M$, take $%
E(\cdot |D,M,X)$ on 
\begin{equation*}
Y=(1-D)(1-M)Y^{00}+(1-D)MY^{01}+D(1-M)Y^{10}+DMY^{11}
\end{equation*}%
and rearrange the resulting conditional means to obtain%
\begin{eqnarray*}
&&E(Y|D,M,X)=E(Y^{00}|D,M,X)\ +\ E(Y^{10}-Y^{00}|D,M,X)\cdot D \\
&&\ \ \ \ \ \ \ +E(Y^{01}-Y^{00}|D,M,X)\cdot M\ +\
E(Y^{11}-Y^{10}-Y^{01}+Y^{00}|D,M,X)\cdot DM \\
&&\ =\mu _{0}(X)+\mu _{1}(X)D+\mu _{4}(X)M+\mu _{3}(X)DM\ \ \ \ \ \text{(due
to C(e)),} \\
&&\ \ \ \ \ \ \ \mu _{0}(X)\equiv E(Y^{00}|X),\ \ \ \ \ \mu _{1}(X)\equiv
E(Y^{10}-Y^{00}|X), \\
&&\ \ \ \ \ \ \ \mu _{4}(X)\equiv E(Y^{01}-Y^{00}|X),\ \ \ \ \mu
_{3}(X)\equiv E(Y^{11}-Y^{10}-Y^{01}+Y^{00}|X)\text{;}
\end{eqnarray*}%
the reason for the subscript $4$ will be seen later. Then $U_{0}\equiv
Y-E(Y|D,M,X)$ gives%
\begin{equation}
Y=\mu _{0}(X)+\mu _{1}(X)D+\mu _{4}(X)M+\mu _{3}(X)DM+U_{0}.  \tag{5.1}
\end{equation}

\qquad We employed two \textquotedblleft linearization
devices\textquotedblright : the uniform error for $M^{d}$ in (4.1) to obtain
the linear model in (4.2), and the approach for the linear-in-$(D,M,DM)$
representation in (5.1). The first subsection below introduces ordinary
least squares estimator (OLS)\ for the former, and the second subsection for
the latter.

\subsection{OLS for Constant Effects}

\qquad From the $M^{d}$ RF in (4.2) and $Y^{dm}$ SF in (4.1), we obtain the
observed variables:%
\begin{eqnarray}
M &=&(1-D)M^{0}+DM^{1}=\alpha _{1}+\alpha _{d}D+X^{\prime }\alpha
_{x}+\varepsilon ,\ \ \ \varepsilon \equiv (1-D)\varepsilon
^{0}+D\varepsilon ^{1},  \notag \\
Y &=&(1-D)(1-M)Y^{00}+(1-D)MY^{01}+D(1-M)Y^{10}+DMY^{11}  \notag \\
&=&\beta _{1}+\beta _{d}D+\beta _{m}M+\beta _{dm}DM+X^{\prime }\beta _{x}+U%
\text{.}  \TCItag{5.2}
\end{eqnarray}%
Since $(\varepsilon ^{0},\varepsilon ^{1})$ are parts of $M^{0}$ and $M^{1}$%
, `$D\amalg (\varepsilon ^{0},\varepsilon ^{1})|X$' holds due to C(a). Hence 
$E(\varepsilon |D,X)=(1-D)E(\varepsilon ^{0}|X)+DE(\varepsilon ^{1}|X)=0$,
and we can obtain%
\begin{equation*}
\text{OLS\ }\hat{\alpha}\text{ of\ }M\text{ on\ }W\equiv (1,D,X^{\prime
})^{\prime }\text{ \ \ for \ \ }\alpha \equiv (\alpha _{1},\alpha
_{d},\alpha _{x}^{\prime })^{\prime }.
\end{equation*}

\qquad Since $Y^{dm}|X$ is determined by $U$, C(e) implies $E(U|D,M,X)=0$,
and we obtain%
\begin{equation*}
\text{OLS\ }\hat{\beta}\text{ of\ }Y\text{ on\ }Z\equiv (1,D,M,DM,X^{\prime
})^{\prime }\text{ \ \ for \ \ }\beta \equiv (\beta _{1},\beta _{d},\beta
_{m},\beta _{dm},\beta _{x}^{\prime })^{\prime }.
\end{equation*}%
Using (4.7), a three-effect decomposition estimator is\ ($\bar{X}$ is the
sample average of $X$)%
\begin{equation*}
\hat{\beta}_{d}+(\hat{\beta}_{m}+\hat{\beta}_{dm})\hat{\alpha}_{d}+\hat{\beta%
}_{dm}(\hat{\alpha}_{1}+\bar{X}^{\prime }\hat{\alpha}_{x}).
\end{equation*}

\qquad For the direct effect estimator $\hat{\beta}_{d}$, with $X$ being of
dimension $k_{x}\times 1$ and $0_{a\times b}$ denoting the null vector of
dimension $a\times b$,%
\begin{eqnarray*}
&&\sqrt{N}(\hat{\beta}_{d}-\beta _{d})\rightarrow ^{d}N(0,\Omega _{1})\text{%
,\ \ \ \ \ }\hat{\Omega}_{1}\equiv \frac{1}{N}\sum_{i}\hat{\eta}%
_{1i}^{2}\rightarrow ^{p}\Omega _{1}, \\
&&\hat{\eta}_{1i}\equiv C_{11}^{\prime }(\frac{1}{N}%
\sum_{i}Z_{i}Z_{i})^{-1}Z_{i}\hat{U}_{i},\ \ \ \hat{U}_{i}\equiv
Y_{i}-Z_{i}^{\prime }\hat{\beta},\ \ \ C_{11}\equiv (0,1,0,0,0_{1\times
k_{x}})^{\prime }.
\end{eqnarray*}%
For the indirect effect estimator $(\hat{\beta}_{m}+\hat{\beta}_{dm})\hat{%
\alpha}_{d}$, the appendix proves%
\begin{eqnarray*}
&&\sqrt{N}\{(\hat{\beta}_{m}+\hat{\beta}_{dm})\hat{\alpha}_{d}-(\beta
_{m}+\beta _{dm})\alpha _{d}\}\rightarrow ^{d}N(0,\Omega _{2})\text{,\ \ \ }%
\hat{\Omega}_{2}\equiv \frac{1}{N}\sum_{i}\hat{\eta}_{2i}^{2}\rightarrow
^{p}\Omega _{2}\text{,} \\
&&\hat{\eta}_{2i}\equiv \hat{C}_{21}^{\prime }(\frac{1}{N}%
\sum_{i}Z_{i}Z_{i})^{-1}Z_{i}\hat{U}_{i}\ +\ \hat{C}_{22}^{\prime }(\frac{1}{%
N}\sum_{i}W_{i}W_{i})^{-1}W_{i}\hat{\varepsilon}_{i}, \\
&&\hat{C}_{21}\equiv (0,0,\hat{\alpha}_{d},\hat{\alpha}_{d},0_{1\times
k_{x}})^{\prime },\ \ \hat{C}_{22}\equiv (0,\hat{\beta}_{m}+\hat{\beta}%
_{dm},0_{1\times k_{x}})^{\prime },\ \ \hat{\varepsilon}_{i}\equiv
M_{i}-W_{i}^{\prime }\hat{\alpha}.
\end{eqnarray*}%
For the interaction effect estimator $\hat{\beta}_{dm}(\hat{\alpha}_{1}+\bar{%
X}^{\prime }\hat{\alpha}_{x})$, the appendix also proves%
\begin{eqnarray}
&&\sqrt{N}[\hat{\beta}_{dm}(\hat{\alpha}_{1}+\bar{X}^{\prime }\hat{\alpha}%
_{x})-\beta _{dm}\{\alpha _{1}+E(X^{\prime })\alpha _{x}\}]\rightarrow
^{d}N(0,\Omega _{3})\text{,\ \ \ }\hat{\Omega}_{3}\equiv \frac{1}{N}\sum_{i}%
\hat{\eta}_{3i}^{2}\rightarrow ^{p}\Omega _{3}\text{,}  \notag \\
&&\hat{\eta}_{3i}\equiv \hat{C}_{31}^{\prime }(\frac{1}{N}%
\sum_{i}Z_{i}Z_{i})^{-1}Z_{i}\hat{U}_{i}\ +\ \hat{C}_{32}^{\prime }(\frac{1}{%
N}\sum_{i}W_{i}W_{i})^{-1}W_{i}\hat{\varepsilon}_{i}\ +\ \hat{\beta}_{dm}%
\hat{\alpha}_{x}^{\prime }(X_{i}-\bar{X}),  \notag \\
&&\hat{C}_{31}\equiv (0,0,0,\hat{\alpha}_{1}+\bar{X}^{\prime }\hat{\alpha}%
_{x},0_{1\times k_{x}})^{\prime },\ \ \ \ \ \hat{C}_{32}\equiv (\hat{\beta}%
_{dm},0,\hat{\beta}_{dm}\bar{X}^{\prime })^{\prime }.  \TCItag{5.3}
\end{eqnarray}%
Finally, for the total effect, its asymptotic variance can be estimated with%
\begin{equation*}
\hat{\Omega}_{123}\equiv \frac{1}{N}\sum_{i}(\hat{\eta}_{1i}+\hat{\eta}_{2i}+%
\hat{\eta}_{3i})^{2}.
\end{equation*}

\subsection{OLS\ for Varying Effects}

\qquad For $X_{0},X_{1},X_{4},X_{3},X_{m}$ consisting of elements of $X$ and
their functions, with $X_{j}$ being of dimension $k_{j}\times 1$, consider
for (5.1) and for indirect effect: 
\begin{eqnarray}
\text{For (5.1)} &:&Y=\beta _{00}^{\prime }X_{0}+\beta _{1x}^{\prime
}X_{1}D+\beta _{4x}^{\prime }X_{4}M+\beta _{3x}^{\prime }X_{3}DM+U_{0}=\beta
_{0}^{\prime }Q_{0}+U_{0},  \notag \\
\text{For indirect} &:&DY=D(\beta _{20}^{\prime }X_{2}+\beta _{2x}^{\prime
}X_{2}M+U_{2})=D(\beta _{2}^{\prime }Q_{2}+U_{2}),  \TCItag{5.4} \\
&&M=\alpha _{m0}^{\prime }X_{m}+\alpha _{mx}^{\prime }X_{m}D+U_{m}=\alpha
_{m}^{\prime }Q_{m}+U_{m},\ \ \alpha _{m}\equiv (\alpha _{m0}^{\prime
},\alpha _{mx}^{\prime })^{\prime },  \notag \\
&&Q_{0}=(X_{0}^{\prime },\ X_{1}^{\prime }D,\ X_{4}^{\prime }M,\
X_{3}^{\prime }DM)^{\prime },\ \ \ \ \ \beta _{0}\equiv (\beta _{00}^{\prime
},\beta _{1x}^{\prime },\beta _{4x}^{\prime },\beta _{3x}^{\prime })^{\prime
},  \notag \\
&&Q_{2}\equiv (X_{2}^{\prime },X_{2}^{\prime }M)^{\prime },\ \ Q_{m}\equiv
(X_{m}^{\prime },X_{m}^{\prime }D)^{\prime },\ \ \beta _{j}\equiv (\beta
_{j0}^{\prime },\beta _{jx}^{\prime })^{\prime },\ \ j=2,m,  \notag
\end{eqnarray}

$U_{j}$, $j=0,2,m$, are error terms, and $DY$ is for the indirect effect
because $E(Y|D=1,M=m)$, $m=0,1$, are needed. The \textquotedblleft
irrelevant\textquotedblright\ subscript $4$ appears in $\beta _{4x}^{\prime
}X_{4}M$, because $\beta _{4x}^{\prime }X_{4}$ is not used in estimating the
three effects.

\qquad The linear models here, which differ much from those in (5.2) based
on (4.1), are to approximate the $X$-conditional intercept and slopes in
(5.1), and other than this, there is no restriction imposed on the data
generating process. That is, $\beta _{00}^{\prime }X_{0}$, $\beta
_{1x}^{\prime }X_{1}$, $\beta _{4x}^{\prime }X_{4}$ and $\beta _{3x}^{\prime
}X_{3}$ are to approximate $\mu _{0}(X)$, $\mu _{1}(X)$, $\mu _{4}(X)$ and $%
\mu _{3}(X)$ in (5.1), and they consist of functions of elements in $X$.
Note that $\beta _{1x}^{\prime }X_{1}=\mu _{1}(X)\equiv E(Y^{10}-Y^{00}|X)$
is the conditional direct effect, $\beta _{3x}^{\prime }X_{3}=\mu
_{3}(X)\equiv E(Y^{11}-Y^{10}-Y^{01}+Y^{00}|X)$ is part of the conditional
interaction effect. However, $\beta _{4x}^{\prime }X_{4}=\mu _{4}(X)\equiv
E(Y^{01}-Y^{00}|X)$ is not the desired indirect effect $%
E\{(Y^{11}-Y^{10})(M^{1}-M^{0})|X\}$, for which the $DY$ model in (5.4)
should be used. The total effect is to be obtained as the sum of the three
effects.

\qquad For varying effects, we condition the inference on $\bar{X}$. This is
not to account for errors of the form $\overline{X_{2}X_{m}^{\prime }}%
-E(X_{2}X_{m}^{\prime })$ relevant for the indirect and interaction effects,
as accounting for such errors requires vectorizing matrices of the form $%
\overline{X_{2}X_{m}^{\prime }}-E(X_{2}X_{m}^{\prime })$, resulting in
unnecessary complications. What is gained by conditioning on $\bar{X}$ is
ease in doing asymptotic inference, as terms like $X_{i}-\bar{X}$ in (5.3)
drop out. What is lost is losing some `external validity', as the findings
apply only to the subpopulation with their $\bar{X}$ values being the same
as those in the sample. Our simulation study will show that not accounting
for errors of the form $\overline{X_{2}X_{m}^{\prime }}-E(X_{2}X_{m}^{\prime
})$ makes little difference.

\qquad For the direct effect $\bar{X}_{1}^{\prime }\beta _{1x}$, doing the
OLS\ of $Y$ on $Q_{0}$, we have%
\begin{eqnarray}
&&\sqrt{N}\bar{X}_{1}^{\prime }(\hat{\beta}_{1x}-\beta _{1x})\rightarrow
^{d}N(0,\Lambda _{1}),\ \ \ \hat{\Lambda}_{1}\equiv \frac{1}{N}\sum_{i}\hat{%
\lambda}_{1i}^{2}\rightarrow ^{p}\Lambda _{1},\ \ \ \hat{U}_{0i}\equiv Y_{i}-%
\hat{\beta}_{0}^{\prime }Q_{0i}  \notag \\
&&G_{1}\equiv (0_{1\times k_{0}},\bar{X}_{1}^{\prime },0_{1\times
(k_{4}+k_{3})})^{\prime },\ \ \ \ \ \hat{\lambda}_{1i}\equiv G_{1}^{\prime }(%
\frac{1}{N}\sum_{i}Q_{0i}Q_{0i}^{\prime })^{-1}Q_{0i}\hat{U}_{0i}. 
\TCItag{5.5}
\end{eqnarray}

\qquad For the indirect effect $E\{(Y^{11}-Y^{10})(M^{1}-M^{0})|X\}$, due to
C(e),%
\begin{equation*}
E\{(Y^{11}-Y^{10})(M^{1}-M^{0})|X\}=E(Y^{11}-Y^{10}|X)\cdot
E(M^{1}-M^{0}|X)=\beta _{2x}^{\prime }X_{2}\cdot X_{m}^{\prime }\alpha _{mx};
\end{equation*}%
$E(Y^{11}-Y^{10}|X)=\beta _{2x}^{\prime }X_{2}$ is obtained from the $DY$
model in (5.4), and $E(M^{1}-M^{0}|X)=X_{m}^{\prime }\alpha _{mx}$ from the $%
M$ model in (5.4). For the product $\beta _{2x}^{\prime }\overline{%
X_{2}X_{m}^{\prime }}\alpha _{mx}$, with $\overline{X_{2}X_{m}^{\prime }}$
being the sample average of $X_{2}X_{m}^{\prime }$, it holds up to an $%
o_{p}(1)$ term that%
\begin{equation*}
\sqrt{N}(\hat{\beta}_{2x}^{\prime }\overline{X_{2}X_{m}^{\prime }}\hat{\alpha%
}_{mx}-\beta _{2x}^{\prime }\overline{X_{2}X_{m}^{\prime }}\alpha
_{mx})=\alpha _{mx}^{\prime }\overline{X_{m}X_{2}^{\prime }}\sqrt{N}(\hat{%
\beta}_{2x}-\beta _{2x})+\beta _{2x}^{\prime }\overline{X_{2}X_{m}^{\prime }}%
\sqrt{N}(\hat{\alpha}_{mx}-\alpha _{mx}).
\end{equation*}%
Then, the appendix proves that%
\begin{eqnarray}
&&\sqrt{N}(\hat{\beta}_{2x}^{\prime }\overline{X_{2}X_{m}^{\prime }}\hat{%
\alpha}_{mx}-\beta _{2x}^{\prime }\overline{X_{2}X_{m}^{\prime }}\alpha
_{mx})\rightarrow ^{d}N(0,\Lambda _{2}),\ \ \ \ \ \hat{\Lambda}_{2}\equiv 
\frac{1}{N}\sum_{i}\hat{\lambda}_{2i}^{2}\rightarrow ^{p}\Lambda _{2}, 
\notag \\
&&\hat{\lambda}_{2i}\equiv \hat{G}_{21}^{\prime }(\frac{1}{N}%
\sum_{i}D_{i}Q_{2i}Q_{2i}^{\prime })^{-1}D_{i}Q_{2i}\hat{U}_{2i}\ +\ \hat{G}%
_{22}^{\prime }(\frac{1}{N}\sum_{i}Q_{mi}Q_{mi}^{\prime })^{-1}Q_{mi}\hat{U}%
_{mi},  \notag \\
&&\hat{G}_{21}\equiv (0_{1\times k_{2}},\hat{\alpha}_{mx}^{\prime }\overline{%
X_{m}X_{2}^{\prime }})^{\prime },\ \ \ \ \ \hat{G}_{22}\equiv (0_{1\times
k_{m}},\hat{\beta}_{2x}^{\prime }\overline{X_{2}X_{m}^{\prime }})^{\prime },
\TCItag{5.6} \\
&&\hat{U}_{2i}\equiv Y_{i}-\hat{\beta}_{2}^{\prime }Q_{2i},\ \ \ \ \ \hat{U}%
_{mi}\equiv M_{i}-\hat{\alpha}_{m}^{\prime }Q_{mi}.  \notag
\end{eqnarray}

\qquad For the interaction effect, due to C(e), we need%
\begin{equation*}
E\{(Y^{11}-Y^{10}-Y^{01}+Y^{00})M^{0}|X\}=\mu _{3}(X)\cdot E(M^{0}|X)=\beta
_{3x}^{\prime }X_{3}\cdot X_{m}^{\prime }\alpha _{m0};
\end{equation*}%
$\beta _{3x}^{\prime }X_{3}$ and $X_{m}^{\prime }\alpha _{m0}$ are obtained
from the $Y$ and $M$ models in (5.4). This gives%
\begin{eqnarray}
&&\sqrt{N}(\hat{\beta}_{3x}^{\prime }\overline{X_{3}X_{m}^{\prime }}\hat{%
\alpha}_{m0}-\beta _{3x}^{\prime }\overline{X_{3}X_{m}^{\prime }}\alpha
_{m0})\rightarrow ^{d}N(0,\Lambda _{3}),\ \ \ \ \ \hat{\Lambda}_{2}\equiv 
\frac{1}{N}\sum_{i}\hat{\lambda}_{3i}^{2}\rightarrow ^{p}\Lambda _{3}, 
\notag \\
&&\hat{\lambda}_{3i}\equiv \hat{G}_{31}^{\prime }(\frac{1}{N}%
\sum_{i}Q_{0i}Q_{0i}^{\prime })^{-1}Q_{0i}\hat{U}_{0i}\ +\ \hat{G}%
_{32}^{\prime }(\frac{1}{N}\sum_{i}Q_{mi}Q_{mi}^{\prime })^{-1}Q_{mi}\hat{U}%
_{mi},  \TCItag{5.7} \\
&&\hat{G}_{31}\equiv (0_{1\times (k_{0}+k_{1}+k_{4})},\hat{\alpha}%
_{m0}^{\prime }\overline{X_{m}X_{3}^{\prime }})^{\prime },\ \ \ \ \ \hat{G}%
_{32}\equiv (\hat{\beta}_{3x}^{\prime }\overline{X_{3}X_{m}^{\prime }}%
,0_{1\times k_{m}})^{\prime }.  \notag
\end{eqnarray}

\qquad The total effect is the sum of the three effects: $\bar{X}%
_{1}^{\prime }\hat{\beta}_{1x}+\hat{\beta}_{2x}^{\prime }\overline{%
X_{2}X_{m}^{\prime }}\hat{\alpha}_{mx}+\hat{\beta}_{3x}^{\prime }\overline{%
X_{3}X_{m}^{\prime }}\hat{\alpha}_{m0}$. The asymptotic variance can be
estimated with $N^{-1}\sum_{i}(\hat{\lambda}_{1i}+\hat{\lambda}_{2i}+\hat{%
\lambda}_{3i})^{2}$.

\section{Simulation and Empirical Analyses}

\subsection{Simulation Study}

\qquad Recalling (4.1), we use four designs in our simulation study with $D$
randomized, $P(D=0)=P(D=1)=0.5$, $N=250,1000$, and $5000$ simulation
repetitions:%
\begin{eqnarray*}
&&\text{Design 1: }M^{d}=1[1<\alpha _{1}+\alpha _{d}d+X^{\prime }\alpha
_{x}+Uni[0,1]]\text{, }X\sim Uni[0,1]\text{, continuous }Y^{dm}; \\
&&\text{Design 2: }M^{d}=1[1<\alpha _{1}+\alpha _{d}d+X^{\prime }\alpha
_{x}+Uni[0,1]]\text{, }X\sim Uni[0,1]\text{, probit }Y^{dm}; \\
&&\text{Design 3: }M^{d}=1[0<\alpha _{1}+\alpha _{d}d+X^{\prime }\alpha
_{x}+N(0,1)]\text{, }X\sim N(0,4)\text{, continuous }Y^{dm}; \\
&&\text{Design 4: }M^{d}=1[0<\alpha _{1}+\alpha _{d}d+X^{\prime }\alpha
_{x}+N(0,1)]\text{, }X\sim N(0,4)\text{, probit }Y^{dm}.
\end{eqnarray*}%
\textquotedblleft Probit $Y^{dm}$\textquotedblright\ means $Y^{dm}=1[0<\ $%
continuous $Y^{dm}]$ with $U\sim N(0,1)$ in $Y^{dm}$. In Designs 1 and 2, $%
E(M^{d}|X)$ is linear. As for the parameter values, we set%
\begin{equation*}
\alpha _{1}=0,\ \alpha _{d}=\alpha _{x}=0.5;\ \beta _{1}=0,\ \beta
_{d}=\beta _{m}=\beta _{dm}=0.5,\ \ \ \beta _{x}=-1;
\end{equation*}%
$\beta _{x}=-1$ is to prevent $Y^{11}$ from having too many zeros. We
generate $M$ and $Y$ with%
\begin{equation*}
M=(1-D)M^{0}+DM^{1},\text{ \ }%
Y=(1-D)(1-M)Y^{00}+(1-D)MY^{01}+D(1-M)Y^{10}+DMY^{11}.
\end{equation*}

\qquad For design 1, the true effects are in (4.6), and for the other
designs, the true effects are found numerically. We use three OLS's: the
constant-effect OLS (\textquotedblleft OLS$_{c}$\textquotedblright ) for
(5.2), the varying effect OLS with $X_{0}=X_{1}=X_{2}=X_{3}=X_{4}=X_{m}=X$
(\textquotedblleft OLS$_{v1}$\textquotedblright ), and the varying effect
OLS\ with $X_{0}=X_{1}=X_{2}=X_{3}=X_{4}=X_{m}$ consisting of $X$ and $X^{2}$
(\textquotedblleft OLS$_{v2}$\textquotedblright ). OLS$_{c}$ is consistent
for Design 1, but there is no guarantee for the consistency of OLS$_{v1}$
and OLS$_{v2}$ for any design, because they approximate unknown functions of 
$X$ as in (5.1) with linear functions of $X$ (and $X^{2}$). Since OLS$_{v2}$
uses more components than OLS$_{v1}$, OLS$_{v2}$ is likely to be less biased
but more variable than OLS$_{v1}$. Only in Design 4, we use
\textquotedblleft OLS$_{v3}$\textquotedblright\ that uses one more component 
$\Phi (X)$ than OLS$_{v2}$ does to improve the approximation.

\qquad Table 1 presents the Design 1 (left half) and Design 2 (right half)
results. Each entry has four numbers: \TEXTsymbol{\vert}Bias\TEXTsymbol{\vert%
}, standard deviation (Sd), root mean squared error (Rmse), and the average
of the $5000$ asymptotic Sd's; the last is to see how accurate the
asymptotic variance formulas are in comparison with the simulation Sd. Since
the effects vary across the designs, we divide all four numbers by the
absolute effect magnitude.

\begin{center}
\begin{tabular}{ccccc}
\hline\hline
\multicolumn{5}{c}{Table 1. \TEXTsymbol{\vert}Bias/effect\TEXTsymbol{\vert},
Sd/\TEXTsymbol{\vert}effect\TEXTsymbol{\vert}, (Rmse/\TEXTsymbol{\vert}effect%
\TEXTsymbol{\vert}) and Asypmtotic-Sd/\TEXTsymbol{\vert}effect\TEXTsymbol{%
\vert}} \\ 
& Design 1, N=250 & Design 1, N=1000 & Design 2, N=250 & Design 2, N=1000 \\ 
\hline
\multicolumn{5}{c}{OLS$_{c}$} \\ 
\multicolumn{1}{l}{tot} & \multicolumn{1}{l}{\small 0.00 0.12 (0.12) 0.12} & 
\multicolumn{1}{l}{\small 0.00 0.06 (0.06) 0.06} & \multicolumn{1}{l}{\small %
0.00 0.14 (0.14) 0.14} & \multicolumn{1}{l}{\small 0.00 0.07 (0.07) 0.07} \\ 
\multicolumn{1}{l}{dir} & \multicolumn{1}{l}{\small 0.01 0.42 (0.42) 0.41} & 
\multicolumn{1}{l}{\small 0.00 0.21 (0.21) 0.21} & \multicolumn{1}{l}{\small %
0.06 0.56 (0.56) 0.54} & \multicolumn{1}{l}{\small 0.05 0.27 (0.28) 0.27} \\ 
\multicolumn{1}{l}{ind} & \multicolumn{1}{l}{\small 0.00 0.24 (0.24) 0.24} & 
\multicolumn{1}{l}{\small 0.00 0.12 (0.12) 0.12} & \multicolumn{1}{l}{\small %
0.03 0.32 (0.32) 0.31} & \multicolumn{1}{l}{\small 0.02 0.15 (0.16) 0.15} \\ 
\multicolumn{1}{l}{int} & \multicolumn{1}{l}{\small 0.01 0.60 (0.60) 0.60} & 
\multicolumn{1}{l}{\small 0.01 0.30 (0.30) 0.30} & \multicolumn{1}{l}{\small %
0.17 0.79 (0.81) 0.77} & \multicolumn{1}{l}{\small 0.13 0.40 (0.42) 0.39} \\ 
\multicolumn{5}{c}{OLS$_{v1}$} \\ 
\multicolumn{1}{l}{tot} & \multicolumn{1}{l}{\small 0.00 0.12 (0.12) 0.12} & 
\multicolumn{1}{l}{\small 0.00 0.06 (0.06) 0.06} & \multicolumn{1}{l}{\small %
0.00 0.14 (0.14) 0.14} & \multicolumn{1}{l}{\small 0.00 0.07 (0.07) 0.08} \\ 
\multicolumn{1}{l}{dir} & \multicolumn{1}{l}{\small 0.01 0.50 (0.50) 0.47} & 
\multicolumn{1}{l}{\small 0.00 0.25 (0.25) 0.24} & \multicolumn{1}{l}{\small %
0.01 0.67 (0.67) 0.63} & \multicolumn{1}{l}{\small 0.01 0.32 (0.32) 0.32} \\ 
\multicolumn{1}{l}{ind} & \multicolumn{1}{l}{\small 0.00 0.28 (0.28) 0.26} & 
\multicolumn{1}{l}{\small 0.00 0.14 (0.14) 0.13} & \multicolumn{1}{l}{\small %
0.00 0.39 (0.39) 0.36} & \multicolumn{1}{l}{\small 0.00 0.18 (0.18) 0.18} \\ 
\multicolumn{1}{l}{int} & \multicolumn{1}{l}{\small 0.01 0.84 (0.84) 0.80} & 
\multicolumn{1}{l}{\small 0.01 0.41 (0.41) 0.40} & \multicolumn{1}{l}{\small %
0.04 1.10 (1.10) 1.05} & \multicolumn{1}{l}{\small 0.01 0.54 (0.54) 0.54} \\ 
\multicolumn{5}{c}{OLS$_{v2}$} \\ 
\multicolumn{1}{l}{tot} & \multicolumn{1}{l}{\small 0.00 0.12 (0.12) 0.12} & 
\multicolumn{1}{l}{\small 0.00 0.06 (0.06) 0.06} & \multicolumn{1}{l}{\small %
0.00 0.14 (0.14) 0.14} & \multicolumn{1}{l}{\small 0.00 0.07 (0.07) 0.08} \\ 
\multicolumn{1}{l}{dir} & \multicolumn{1}{l}{\small 0.01 0.64 (0.64) 0.51} & 
\multicolumn{1}{l}{\small 0.00 0.27 (0.27) 0.26} & \multicolumn{1}{l}{\small %
0.01 0.85 (0.85) 0.68} & \multicolumn{1}{l}{\small 0.01 0.35 (0.35) 0.34} \\ 
\multicolumn{1}{l}{ind} & \multicolumn{1}{l}{\small 0.00 0.36 (0.36) 0.29} & 
\multicolumn{1}{l}{\small 0.00 0.15 (0.15) 0.14} & \multicolumn{1}{l}{\small %
0.00 0.51 (0.51) 0.40} & \multicolumn{1}{l}{\small 0.01 0.20 (0.20) 0.19} \\ 
\multicolumn{1}{l}{int} & \multicolumn{1}{l}{\small 0.01 1.13 (1.13) 0.91} & 
\multicolumn{1}{l}{\small 0.01 0.47 (0.47) 0.45} & \multicolumn{1}{l}{\small %
0.02 1.46 (1.46) 1.20} & \multicolumn{1}{l}{\small 0.03 0.62 (0.63) 0.60} \\ 
\multicolumn{1}{l}{tru} & \multicolumn{2}{c}{1.125, 0.500, 0.500, 0.125} & 
\multicolumn{2}{c}{0.395, 0.184, 0.166, 0.045} \\ \hline
\multicolumn{5}{c}{{\small OLS}$_{c}${\small \ for constant-effect (5.2); OLS%
}$_{v1}${\small \ \& OLS}$_{v2}${\small \ for }${\small X}${\small %
-heterogeneous\ effect approximations;}} \\ 
\multicolumn{5}{c}{\small tot for total effect; dir for direct; ind for
indirect; int for interaction; tru for true effect} \\ \hline\hline
\end{tabular}
\end{center}

\qquad In Design 1 with $N=250$, all biases are almost zero, and OLS$_{c}$
does best, followed by OLS$_{v1}$ and then OLS$_{v2}$ that is more variable
than OLS$_{v1}$. With $N=1000$, all OLS's improve, and the performance
differences narrow. In Design 2 with binary $Y$, although OLS$_{c}$ still
does the best followed by OLS$_{v1}$ and OLS$_{v2}$, OLS$_{c}$ is biased,
particularly for the interaction effect, and the biases decrease little even
when $N$ increases to $1000$. Although omitted from Table 1, due to the
bias, OLS$_{c}$ is dominated eventually, as $N$ increases beyond $1000$. The
second and fourth numbers in each entry of Table 1 are mostly the same,
showing that the asymptotic variance formulas are accurate. For this, not
accounting for the errors of the form $\bar{X}-E(X)$ in OLS$_{v1}$ and OLS$%
_{v2}$ hardly matters.

\begin{center}
\begin{tabular}{ccccc}
\hline\hline
\multicolumn{5}{c}{Table 2. \TEXTsymbol{\vert}Bias/effect\TEXTsymbol{\vert},
Sd/effect, (Rmse/effect) and Asypmtotic-Sd/effect} \\ 
& Design 3, N=250 & Design 3, N=1000 & Design 4, N=250 & Design 4, N=1000 \\ 
\hline
\multicolumn{5}{c}{OLS$_{c}$} \\ 
\multicolumn{1}{l}{tot} & \multicolumn{1}{l}{\small 0.00 0.15 (0.15) 0.15} & 
\multicolumn{1}{l}{\small 0.00 0.07 (0.07) 0.07} & \multicolumn{1}{l}{\small %
0.01 0.27 (0.27) 0.26} & \multicolumn{1}{l}{\small 0.00 0.13 (0.13) 0.13} \\ 
\multicolumn{1}{l}{dir} & \multicolumn{1}{l}{\small 0.01 0.39 (0.39) 0.39} & 
\multicolumn{1}{l}{\small 0.01 0.20 (0.20) 0.20} & \multicolumn{1}{l}{\small %
0.18 0.73 (0.75) 0.72} & \multicolumn{1}{l}{\small 0.16 0.36 (0.40) 0.36} \\ 
\multicolumn{1}{l}{ind} & \multicolumn{1}{l}{\small 0.01 0.44 (0.44) 0.43} & 
\multicolumn{1}{l}{\small 0.01 0.21 (0.21) 0.21} & \multicolumn{1}{l}{\small %
0.10 0.60 (0.61) 0.59} & \multicolumn{1}{l}{\small 0.09 0.29 (0.30) 0.29} \\ 
\multicolumn{1}{l}{int} & \multicolumn{1}{l}{\small 0.00 0.53 (0.53) 0.52} & 
\multicolumn{1}{l}{\small 0.01 0.26 (0.26) 0.26} & \multicolumn{1}{l}{\small %
0.24 0.81 (0.85) 0.81} & \multicolumn{1}{l}{\small 0.23 0.41 (0.47) 0.40} \\ 
\multicolumn{5}{c}{OLS$_{v1}$} \\ 
\multicolumn{1}{l}{tot} & \multicolumn{1}{l}{\small 0.00 0.15 (0.15) 0.15} & 
\multicolumn{1}{l}{\small 0.00 0.07 (0.07) 0.07} & \multicolumn{1}{l}{\small %
0.01 0.27 (0.27) 0.26} & \multicolumn{1}{l}{\small 0.00 0.13 (0.13) 0.13} \\ 
\multicolumn{1}{l}{dir} & \multicolumn{1}{l}{\small 0.01 0.52 (0.52) 0.50} & 
\multicolumn{1}{l}{\small 0.00 0.26 (0.26) 0.25} & \multicolumn{1}{l}{\small %
0.71 1.20 (1.40) 1.17} & \multicolumn{1}{l}{\small 0.73 0.59 (0.94) 0.59} \\ 
\multicolumn{1}{l}{ind} & \multicolumn{1}{l}{\small 0.01 0.46 (0.46) 0.46} & 
\multicolumn{1}{l}{\small 0.01 0.22 (0.22) 0.22} & \multicolumn{1}{l}{\small %
0.24 0.67 (0.71) 0.67} & \multicolumn{1}{l}{\small 0.21 0.31 (0.37) 0.32} \\ 
\multicolumn{1}{l}{int} & \multicolumn{1}{l}{\small 0.01 0.83 (0.83) 0.79} & 
\multicolumn{1}{l}{\small 0.00 0.40 (0.40) 0.39} & \multicolumn{1}{l}{\small %
1.05 1.49 (1.82) 1.45} & \multicolumn{1}{l}{\small 1.07 0.73 (1.30) 0.72} \\ 
\multicolumn{5}{c}{OLS$_{v2}$} \\ 
\multicolumn{1}{l}{tot} & \multicolumn{1}{l}{\small 0.00 0.15 (0.15) 0.15} & 
\multicolumn{1}{l}{\small 0.00 0.07 (0.07) 0.07} & \multicolumn{1}{l}{\small %
0.02 0.26 (0.26) 0.25} & \multicolumn{1}{l}{\small 0.00 0.13 (0.13) 0.13} \\ 
\multicolumn{1}{l}{dir} & \multicolumn{1}{l}{\small 0.02 0.71 (0.72) 0.62} & 
\multicolumn{1}{l}{\small 0.00 0.32 (0.32) 0.31} & \multicolumn{1}{l}{\small %
0.12 1.60 (1.61) 1.36} & \multicolumn{1}{l}{\small 0.15 0.72 (0.73) 0.68} \\ 
\multicolumn{1}{l}{ind} & \multicolumn{1}{l}{\small 0.01 0.49 (0.49) 0.49} & 
\multicolumn{1}{l}{\small 0.01 0.22 (0.22) 0.23} & \multicolumn{1}{l}{\small %
0.03 0.81 (0.81) 0.77} & \multicolumn{1}{l}{\small 0.01 0.34 (0.34) 0.34} \\ 
\multicolumn{1}{l}{int} & \multicolumn{1}{l}{\small 0.02 1.24 (1.24) 1.07} & 
\multicolumn{1}{l}{\small 0.00 0.55 (0.55) 0.52} & \multicolumn{1}{l}{\small %
0.22 2.17 (2.18) 1.82} & \multicolumn{1}{l}{\small 0.23 0.97 (0.99) 0.91} \\ 
\multicolumn{5}{c}{OLS$_{v3}$} \\ 
tot &  &  & {\small 0.01 0.25 (0.25) 0.25} & {\small 0.01 0.13 (0.13) 0.12}
\\ 
dir &  &  & {\small 0.09 1.98 (1.98) 1.35} & {\small 0.04 0.55 (0.55) 0.52}
\\ 
ind &  &  & {\small 0.07 0.87 (0.87) 0.81} & {\small 0.01 0.31 (0.31) 0.32}
\\ 
int &  &  & {\small 0.12 2.93 (2.93) 1.98} & {\small 0.07 0.77 (0.78) 0.71}
\\ 
\multicolumn{1}{l}{tru} & \multicolumn{2}{c}{0.888, 0.500, 0.138, 0.250} & 
\multicolumn{2}{c}{0.169, 0.089, 0.024, 0.055} \\ \hline
\multicolumn{5}{c}{{\small OLS}$_{c}${\small \ for constant-effect (5.2); OLS%
}$_{v1}${\small ,OLS}$_{v2}$,{\small OLS}$_{v3}${\small \ for }${\small X}$%
{\small -heterogeneous\ effect approximations;}} \\ 
\multicolumn{5}{c}{\small tot for total effect; dir for direct; ind for
indirect; int for interaction; tru for true effect} \\ \hline\hline
\end{tabular}
\end{center}

\qquad Table 2 presents the Design 3 (left half) and Design 4 (right half)
results. In Design 3 with continuous $Y$, OLS$_{c}$ does best followed by OLS%
$_{v1}$ and OLS$_{v2}$ as in Table 1, showing that, when $Y$ is continuous,
the binary model with uniform error in (4.1) is not a bad specification. In
Design 4 with binary $Y$, however, OLS$_{c}$ has large biases that do not
decrease even when $N$ goes up. Hence it is hard to recommend OLS$_{c}$
although it is still the best in terms of Rmse. OLS$_{v1}$ is even more
biased, which is also hard to recommend. OLS$_{v2}$ has the smallest biases,
which, however, do not drop as $N$ increases. Since OLS$_{c}$, OLS$_{v1}$
and OLS$_{v2}$ do poorly in terms of bias in Design 4, to see if bias can be
reduced further, we use $\Phi (X)$ in addition to $X$ and $X^{2}$ in $%
X_{0}=X_{1}=X_{2}=X_{3}=X_{4}=X_{m}$ to get OLS$_{v3}$. Indeed, OLS$_{v3}$
has biases much smaller than the other OLS's.

\qquad In summary, first, the OLS for the constant effect model (5.2) does
surprisingly well, despite its uniform error specification for $M$. Second,
the OLS's approximating unknown heterogeneous effects with linear functions
do slightly worse, but their performances catch up as $N$ goes up. Third,
the OLS's using more extensive specifications to approximate unknown
heterogenous effects tend to be more variable, but they are well worth
trying due to the lower biases. Fourth, the asymptotic Sd formulas of our
estimators work well, and not accounting for the errors of the form $\bar{X}%
-E(X)$ in the varying-effect estimators hardly matters.

\subsection{Empirical Analysis}

\qquad Our empirical analysis uses the National Longitudinal Survey data in
Card (1995), which are downloadable from
`http://davidcard.berkeley.edu/data\_sets.html' as of this writing; the data
have been used also in Tan (2010) and Wang et al. (2017), among others. In
our empirical analysis with $N=3010$, $Y$ is $\ln ($wage in 1976$)$, $D$ is
the dummy for black, $M$ is the dummy for college education (i.e., schooling
years being 12 or greater), and $X$ consists of age, dummies
(\textquotedblleft r1, r2, ...\textquotedblright ) for 8 residence regions
in 1966, dummy for living in a standard metropolitan statistical area (SMSA)
in 1966, dummy for living in SMSA in 1976 (\textquotedblleft SMSA$_{76}$%
\textquotedblright ), and dummy for living in South in 1976
(\textquotedblleft south\textquotedblright ). In the original data, there
were 9 residence region dummies, but the dummy for region 8 was dropped due
to a singularity problem in our OLS's. That is, we set%
\begin{equation*}
X=(\text{1,\ age,\ r1,\ r2,\ r3,\ r4,\ r5,\ r6,\ r7,\ r9,\ SMSA,\ SMSA}_{76}%
\text{,\ South})^{\prime }.
\end{equation*}

\qquad The data set is old, but this suits well our purpose of finding
racial discrimination effect on wage, which consists of the direct effect,
the indirect effect through college education, and the interaction effect of
black and college education. When gender discrimination cases were argued in
court in the past, often the counter-argument was that females were less
educated/qualified, but lower education/qualification itself might have been
due to gender discrimination. Hence it is important to account for the
indirect discrimination through missed education opportunities, but doing so
with recent data would be difficult because discrimination due to denied
education opportunities is unlikely to be present. For this reason, using an
old data set as ours is advantageous.

\qquad Table 3 presents the estimation results, where OLS$_{c}$ and OLS$%
_{v1} $ are the same as those in the simulation study, but OLS$_{v2}$ is
different because only age is continuous in $X$ with all the other
covariates being binary. For OLS$_{v2}$, we use additionally all interaction
terms between age and the other components of $X$ for $%
X_{0}=X_{1}=X_{2}=X_{3}=X_{4}=X_{m}$.

\begin{center}
\begin{tabular}{lccc}
\hline\hline
\multicolumn{4}{c}{Table 3. Effects of Black with College Education as
Mediator} \\ 
& OLS$_{c}$ (t-value) & OLS$_{v1}$ (t-value) & OLS$_{v2}$ (t-value) \\ \hline
total effect & -0.243 (-13) & -0.223 (-9.0) & -0.220 (-8.5) \\ 
direct effect & -0.272 (-12) & -0.242 (-7.7) & -0.248 (-8.0) \\ 
indirect effect & -0.054 (-6.3) & -0.075 (-4.9) & -0.075 (-4.5) \\ 
interaction effect & 0.083 (4.4) & 0.094 (3.9) & 0.103 (4.3) \\ \hline
\multicolumn{4}{l}{OLS$_{c}$ for the constant-effect model (5.2) with above $%
X$} \\ 
\multicolumn{4}{l}{OLS$_{v1}$ approximates $X$-heterogeneous effects with
linear functions of $X$} \\ 
\multicolumn{4}{l}{OLS$_{v2}$ additionally uses the interactions between age
and the other $X$ elements} \\ \hline\hline
\end{tabular}
\end{center}

\qquad Regardless of the estimator in use, the estimates are similar, and
all effects are statistically significant. The total effect of being black
on wage is about $-22\sim -24\%$, which consists of the direct effect $%
-24\sim -27\%$, the indirect effect of $-5.4\sim -7.5\%$ through missed
college education opportunities, and the interaction effect $8.3\sim 10\%$.
That is, had it not been for the indirect effect through college education,
the wage discrimination would have been lesser by $-5.4\sim -7.5\%$, and
college education alleviated the racial discrimination by $8.3\sim 10\%$.

\section{Conclusions}

\qquad A treatment $D$ can affect an outcome $Y$ indirectly through a
mediator $M$, as well as directly. $D$ can also interact with $M$ to affect $%
Y$. In the literature of mediation analysis decomposing the total effect,
this interaction effect has been part of the indirect effect, which seems
however inappropriate, because $D$ and $M$ in the interaction term $DM$ are
on an equal footing, differently from the indirect effect where $D$ precedes 
$M$.

\qquad In this paper, we proposed decomposing the total effect into three
effects: direct, indirect and interaction effects. In addition to the
advantage of separating the interaction effect from the indirect effect, our
decomposition is \textquotedblleft path-free\textquotedblright , in the
sense that there is no other sensible way to carry out three-effect
decomposition. This is in contrast to the existing path-dependent two-way
(direct and indirect) decompositions.

\qquad After presenting our three-way decomposition, we showed how to
identify them, which was then followed by two OLS-based estimators. The
first OLS assumes an uniform-distributed error for $M$, which essentially
linearizes $E(M|X)$ for covariates $X$. The second OLS\ does not make such
an assumption; instead, it establishes a linear-in-$(D,M,DM)$ representation
for almost any form of $Y$, with $(D,M,DM)$ carrying $X$-heterogeneous
slopes/effects. The second OLS then approximates the unknown $X$%
-heterogeneous effects with linear functions of $X$.

\qquad We carried out a simulation study to demonstrate that the two OLS's
work well. Then we applied the estimators to a data set with $D$ being the
dummy for black, $M$ being college education dummy, and $Y=\ln ($wage$)$. We
found out that the total effect of black dummy on wage is about $-23\%$
consisting of $-25\%$ (direct effect), $-7\%$ (indirect effect through
missed college education opportunities), and $9\%$ (interaction
effect).\medskip \bigskip 

\begin{center}
{\LARGE APPENDIX}
\end{center}

\textbf{Proof for (4.5) and (4.6)}\bigskip

\qquad Observe%
\begin{eqnarray*}
&&\ \delta
(d)=Y^{1,M^{d}}-Y^{0,M^{d}}=Y^{10}+(Y^{11}-Y^{10})M^{d}-%
\{Y^{00}+(Y^{01}-Y^{00})M^{d}\} \\
&&\ =Y^{10}-Y^{00}+(Y^{11}-Y^{10}-Y^{01}+Y^{00})M^{d}=\beta _{d}+\beta
_{dm}M^{d}; \\
&&\mu
(d)=Y^{d,M^{1}}-Y^{d,M^{0}}=Y^{d0}+(Y^{d1}-Y^{d0})M^{1}-%
\{Y^{d0}+(Y^{d1}-Y^{d0})M^{0}\} \\
&&\ =(Y^{d1}-Y^{d0})(M^{1}-M^{0})=(\beta _{d}d+\beta _{m}+\beta _{dm}d-\beta
_{d}d)(M^{1}-M^{0}) \\
&&\ =\beta _{m}(M^{1}-M^{0})+\beta _{dm}d(M^{1}-M^{0}).
\end{eqnarray*}%
Since $M^{1}-M^{0}=\alpha _{d}+\varepsilon ^{1}-\varepsilon ^{0}$ from
(4.2), (1.4) is%
\begin{eqnarray*}
&&\mu (1)+\delta (0)=(\beta _{m}+\beta _{dm})(M^{1}-M^{0})+\beta _{d}+\beta
_{dm}M^{0}=\beta _{d}+\beta _{m}(M^{1}-M^{0})+\beta _{dm}M^{1} \\
&&\ =\beta _{d}\ +\ \beta _{m}(\alpha _{d}+\varepsilon ^{1}-\varepsilon
^{0})\ +\ \beta _{dm}(\alpha _{1}+\alpha _{d}+X^{\prime }\alpha
_{x}+\varepsilon ^{1}).
\end{eqnarray*}%
Taking $E(\cdot )$ removes $\varepsilon ^{1}$ and $\varepsilon ^{0}$, and
gives $E(\tau )$ in (4.6).\bigskip

\textbf{Proof for Indirect-Effect Distribution for Constant-Effect
Model\bigskip }

\qquad With $E^{-1}(\cdot )\equiv \{E(\cdot )\}^{-1}$, it holds up to $%
o_{p}(1)$ terms that%
\begin{eqnarray*}
&&\sqrt{N}(\hat{\alpha}-\alpha )=\frac{1}{\sqrt{N}}\sum_{i}E^{-1}(WW^{\prime
})W_{i}\varepsilon _{i},\ \ \ \sqrt{N}(\hat{\beta}-\beta )=\frac{1}{\sqrt{N}}%
\sum_{i}E^{-1}(ZZ^{\prime })Z_{i}U_{i}; \\
&&\sqrt{N}\{(\hat{\beta}_{m}+\hat{\beta}_{dm})\hat{\alpha}_{d}-(\beta
_{m}+\beta _{dm})\alpha _{d}\} \\
&&\ =\alpha _{d}\sqrt{N}(\hat{\beta}_{m}-\beta _{m})+\alpha _{d}\sqrt{N}(%
\hat{\beta}_{dm}-\beta _{dm})+(\beta _{m}+\beta _{dm})\sqrt{N}(\hat{\alpha}%
_{d}-\alpha _{d}) \\
&&\ =C_{21}^{\prime }\sqrt{N}(\hat{\beta}-\beta )+C_{22}^{\prime }\sqrt{N}(%
\hat{\alpha}-\alpha ) \\
&&\ =\frac{1}{\sqrt{N}}\sum_{i}\{C_{21}^{\prime }E^{-1}(ZZ^{\prime
})Z_{i}U_{i}+C_{22}^{\prime }E^{-1}(WW^{\prime })W_{i}\varepsilon _{i}\} \\
&&\text{where \ \ }C_{21}\equiv (0,0,\alpha _{d},\alpha _{d},0_{1\times
k_{x}})^{\prime }\text{ \ \ \ \ and \ \ \ \ }C_{22}\equiv (0,\beta
_{m}+\beta _{dm},0_{1\times k_{x}})^{\prime }.
\end{eqnarray*}

Defining $\eta _{2i}\equiv C_{21}^{\prime }E^{-1}(ZZ^{\prime
})Z_{i}U_{i}+C_{22}^{\prime }E^{-1}(WW^{\prime })W_{i}\varepsilon _{i}$,
this is asymptotically normal with variance $\Omega _{2}\equiv E(\eta
_{2}\eta _{2}^{\prime })$.\bigskip

\textbf{Proof for (5.3)\bigskip }

\qquad With $\xi \equiv E(X)$, it holds up to $o_{p}(1)$ terms that%
\begin{eqnarray*}
&&\sqrt{N}\{\hat{\beta}_{dm}(\hat{\alpha}_{1}+\bar{X}^{\prime }\hat{\alpha}%
_{x})-\beta _{dm}(\alpha _{1}+\xi ^{\prime }\alpha _{x})\}=(\alpha _{1}+\xi
^{\prime }\alpha _{x})\sqrt{N}(\hat{\beta}_{dm}-\beta _{dm}) \\
&&\ \ \ \ \ \ \ \ \ \ \ \ \ \ \ \ \ +\beta _{dm}\sqrt{N}(\hat{\alpha}%
_{1}-\alpha _{1})+\beta _{dm}\xi ^{\prime }\sqrt{N}(\hat{\alpha}_{x}-\alpha
_{x})+\beta _{dm}\alpha _{x}^{\prime }(\bar{X}-\xi ) \\
&&\ =\frac{1}{\sqrt{N}}\sum_{i}\{C_{31}^{\prime }E^{-1}(ZZ^{\prime
})Z_{i}U_{i}+C_{32}^{\prime }E^{-1}(WW^{\prime })W_{i}\varepsilon _{i}+\beta
_{dm}\alpha _{x}^{\prime }(X_{i}-\xi )\}, \\
&&\text{where \ \ \ \ }C_{31}\equiv (0,0,0,\alpha _{1}+\xi ^{\prime }\alpha
_{x},0_{1\times k_{x}})^{\prime },\ \ \ \ \ C_{32}\equiv (\beta
_{dm},0,\beta _{dm}\xi ^{\prime })^{\prime }.
\end{eqnarray*}%
Defining $\eta _{3i}\equiv C_{31}^{\prime }E^{-1}(ZZ^{\prime
})Z_{i}U_{i}+C_{32}^{\prime }E^{-1}(WW^{\prime })W_{i}\varepsilon _{i}+\beta
_{dm}\alpha _{x}^{\prime }(X_{i}-\xi )$, this is asymptotically normal with
variance $\Omega _{3}\equiv E(\eta _{3}\eta _{3}^{\prime })$.\bigskip

\textbf{Proof for (5.6)\bigskip }

\qquad Given $D=1$, we have $Y=(Y^{11}-Y^{10})M+Y^{10}$. Take $E(\cdot
|D=1,M,X)$ on this $Y$:%
\begin{eqnarray*}
&&E(Y|D=1,M,X)=E(Y^{11}-Y^{10}|D=1,M,X)M+E(Y^{10}|D=1,M,X) \\
&&\ =E(Y^{11}-Y^{10}|X)M+E(Y^{10}|X)\ \ \ \ \ \text{(due to C(e)).}
\end{eqnarray*}%
Substitute $U_{2}\equiv Y-E(Y|D=1,M,X)$ into this $E(Y|D=1,M,X)$ model:%
\begin{eqnarray*}
&&Y=\mu _{20}(X)+\mu _{21}(X)M+U_{2}\text{,}\ \ \ \ \ E(U_{2}|D=1,M,X)=0, \\
&&\mu _{20}(X)=E(Y^{10}|X)\text{ \ \ \ \ and \ \ \ \ }\mu
_{21}(X)=E(Y^{11}-Y^{10}|X)\text{.}
\end{eqnarray*}%
With $\mu _{20}(X)=\beta _{20}^{\prime }X_{2}$ and $\mu
_{21}(X)=E(Y^{11}-Y^{10}|X)=\beta _{2x}^{\prime }X_{2}$, we obtain%
\begin{equation*}
DY=D(\beta _{20}^{\prime }X_{2}+\beta _{2x}^{\prime
}X_{2}M+U_{2})=D(Q_{2}^{\prime }\beta _{2}+U_{2}).
\end{equation*}%
The estimand of the OLS\ of $DY$ on $DQ_{2}$ is%
\begin{eqnarray*}
&&E^{-1}(DQ_{2}Q_{2}^{\prime })E(DQ_{2}Y)=E^{-1}(DQ_{2}Q_{2}^{\prime
})E\{DQ_{2}(Q_{2}^{\prime }\beta _{2}+U_{2})\}=\beta _{2}, \\
&&\text{as }E(DQ_{2}U_{2})=E[\ E\{Q_{2}E(U_{2}|D=1,M,X)|D=1,M,X\}\cdot
P(D=1|M,X)\ ]=0.
\end{eqnarray*}%
\medskip

\begin{center}
{\LARGE REFERENCES}
\end{center}

\qquad Abadie, A. and G. Imbens, 2016, Matching on the estimated propensity
score, Econometrica 84, 781-807.

\qquad Card, D., 1995, Using geographic variation in college proximity to
estimate the return to schooling, in Aspects of Labor Market Behavior:
Essays in Honour of John Vanderkamp, edited by L. Christofides, E. Grant and
R. Swidinsky, 201-222, University of Toronto Press, Toronto.

\qquad Diaz, I. and N.S. Hejazi, 2020, Causal mediation analysis for
stochastic interventions, Journal of the Royal Statistical Society (Series
B) 82, 661-683.

\qquad Diaz, I., N.S. Hejazi, K.E. Rudolph and M.J. van der Laan, 2021,
Nonparametric efficient causal mediation with intermediate confounders,
Biometrika 108, 627-641.

\qquad Ding, P. and J. Lu, 2017, Principal stratification analysis using
principal scores, Journal of the Royal Statistical Society (Series B) 79,
757-777.

\qquad Forastiere, L., A. Mattei and P. Ding, 2018, Principal ignorability
in mediation analysis: through and beyond sequential ignorability,
Biometrika 105, 979-986.

\qquad Fr\"{o}lich, M. and M. Huber, 2017, Direct and indirect treatment
effects--causal chains and mediation analysis with instrumental variables,
Journal of the Royal Statistical Society (Series B) 79, 1645-1666.

\qquad Imai, K., L. Keele and T. Yamamoto, 2010, Identification, inference,
and sensitivity analysis for causal mediation effects, Statistical Science
25, 51-71.

\qquad Imbens, G.W. and J.D. Angrist, 1994, Identification and estimation of
local average treatment effects, Econometrica 62, 467-475.

\qquad Jo, B. and E.A. Stuart, 2009, On the use of propensity scores in
principal causal effect estimation, Statistics in Medicine 28, 2857-2875.

\qquad Joffe, M.M., D. Small, T. Ten Have, S. Brunelli and H.I. Feldman,
2008, Extended instrumental variables estimation for overall effects,
International Journal of Biostatistics 4, 1-20.

\qquad Lee, M.J., 2012, Treatment effects in sample selection models and
their nonparametric estimation, Journal of Econometrics 167, 317-329.

\qquad Lee, M.J., 2017, Extensive and intensive margin effects in sample
selection models: racial effects on wage,\ Journal of the Royal Statistical
Society (Series A) 180, 817-839.

\qquad Lee, M.J. 2018, Simple least squares estimator for treatment effects
using propensity score residuals, Biometrika 105, 149-164.

\qquad Lee, M.J., 2021, Instrument residual estimator for any response
variable with endogenous binary treatment, Journal of the Royal Statistical
Society (Series B) 83, 612-635.

\qquad Lok, J.J., 2016, Defining and estimating causal direct and indirect
effects when setting the mediator to specific values is not feasible,
Statistics in Medicine 35, 4008-4020.

\qquad MacKinnon, D.P., A.J. Fairchild and M.S. Fritz, 2007, Mediation
analysis, Annual Review of Psychology 58, 593-614.

\qquad Nguyen, T.Q., I. Schmid and E.A. Stuart, 2021, Clarifying causal
mediation analysis for the applied researcher: defining effects based on
what we want to learn, Psychological Methods 26, 255-271.

\qquad Pearl, J., 2001, Direct and indirect effects, in Proceedings of the
Seventeenth Conference on Uncertainty in Artificial Intelligence, San
Francisco, CA, Morgan Kaufman, pp. 411-420.

\qquad Pearl, J., 2009, Causality, 2nd ed., Cambridge University Press.

\qquad Petersen, M.L., S.E. Sinisi and M.J. van der Laan, 2006, Estimation
of direct causal effects, Epidemiology 17, 276-284.

\qquad Preacher, K.J. 2015, Advances in mediation analysis: a survey and
synthesis of new developments, Annual Review of Psychology 66, 825-852.

\qquad Robins, J.M., 2003, Semantics of causal DAG models and the
identification of direct and indirect effects, In Highly Structured
Stochastic Systems, edited by P.J. Green, N.L. Hjort and S. Richardson,
70-81, Oxford University Press, Oxford.

\qquad Rubin, D.B., 2004, Direct and indirect causal effects via potential
outcomes, Scandinavian Journal of Statistics 31, 161-170.

\qquad Rudolph, K.E., O. Sofrygin and Mark J. van der Laan, 2021, Complier
stochastic direct effects: identification and robust estimation, Journal of
the American Statistical Association, forthcoming.

\qquad Tan, Z., 2010, Marginal and nested structural models using
instrumental variables, Journal of the American Statistical Association 105,
157-169.

\qquad Tchetgen Tchetgen, E.J. and I. Shpitser, 2012, Semiparametric theory
for causal mediation analysis: efficiency bounds, multiple robustness and
sensitivity analysis, Annals of Statistics 40, 1816-1845.

\qquad Tchetgen Tchetgen, E.J. and I. Shpitser, 2014, Estimation of a
semiparametric natural direct effect model incorporating baseline
covariates, Biometrika 101, 849-864.

\qquad TenHave, T.R. and M.M. Joffe, 2012, A review of causal estimation of
effects in mediation analyses, Statistical Methods in Medical Research 21,
77-107.

\qquad VanderWeele, T.J., 2015, Explanation in causal inference: methods for
mediation and interaction, Oxford University Press.

\qquad VanderWeele, T.J., 2016, Mediation analysis:\ a practitioner's guide,
Annual Review of Public Health 37, 17-32.

\qquad VanderWeele, T.J. and E.J. Tchetgen Tchetgen, 2017, Mediation
analysis with time varying exposures and mediators, Journal of the Royal
Statistical Society 79 (Series B), 917-938.

\qquad VanderWeele, T.J., S. Vansteelandt and J.M. Robins, 2014, Effect
decomposition in the presence of an exposure-induced mediator-outcome
confounder, Epidemiology 25, 300-306.

\qquad Vansteelandt, S. and R.M. Daniel, 2014, On regression adjustment for
the propensity score, Statistics in Medicine 33, 4053-4072.

\qquad Vansteelandt, S. and R.M. Daniel, 2017, Interventional effects for
mediation analysis with multiple mediators, Epidemiology 28, 258-265.

\qquad Wang, L., J.M. Robins and T.S. Richardson, 2017, On falsification of
the binary instrumental variable model, Biometrika 104, 229-236.

\end{document}